\documentclass[english]{lipics-v2021}

\usepackage[T1]{fontenc}
\usepackage[utf8]{inputenc}
\usepackage{microtype}
\usepackage{amsmath,amssymb,amsthm}
\usepackage{algorithm}
\usepackage[noend]{algpseudocode}
\usepackage[normalem]{ulem}
\usepackage{todonotes}

\makeatletter
\@ifclasswith{lipics-v2021}{anonymous}{\let\ifanonymous\iftrue}{\let\ifanonymous\iffalse}
\makeatother

\newenvironment{proofof}[1]{ \noindent {\bf Proof of #1.}\quad }{\qed\par\vskip 4mm\par}
\numberwithin{equation}{section}
\newcommand{\STC}{\operatorname{STC}}
\newcommand{\graph}{F_{\CP}}

\newcommand{\Ot}{\widetilde O}

\newcommand{\CP}{\mathcal P}
\newcommand{\CC}{\mathcal C}
\newcommand{\CL}{\mathcal L}
\newcommand{\CS}{\mathcal S} 
\newcommand{\calP}{\mathcal P}

\newcommand{\NP}{\mathsf{NP}}
\newcommand{\PP}{\mathsf{P}}

\newcommand{\mincut}{\operatorname{mincut}}
\newcommand{\isolalg}{\textsc{IsolatingSets}}
\newcommand{\lamalg}{\textsc{LaminarConstruction}}
\newcommand{\Chandranalg}{\textsc{ChandranCheungIssacTree}}

\newcommand{\oldvalue}[1]{\marginpar{\raggedright\scriptsize\color{red}\normalfont Old: #1}}
\newcommand{\margincomment}[1]%
{\marginpar{\tiny{\color{red}#1}}}
\newcommand{\petr}[1]%
{\marginpar{\tiny{\color{blue}#1}}}
\ifanonymous
\renewcommand{\oldvalue}[1]{}
\fi

\title{Sublinear Approximations for Spanning Tree Congestion}

\titlerunning{Sublinear Approximations for Spanning Tree Congestion}

\author{Sunny Atalig}{University of California at Riverside, USA}{sunny.atalig@ucr.edu}{https://orcid.org/0009-0007-2290-152X}{}

\author{Marek Chrobak}{University of California at Riverside, USA \and \url{https://www.cs.ucr.edu/~marek/}}{marek@cs.ucr.edu}{https://orcid.org/0000-0002-8673-2709}{Partially supported by NFS grant CCF-2153723.}

\author{Petr Kolman}{Charles University, Faculty of Mathematics and Physics, Department of Applied Mathematics, Czechia \and \url{https://kam.mff.cuni.cz/~kolman/}}{kolman@kam.mff.cuni.cz}{https://orcid.org/0000-0003-2235-0506}{Partially supported by GA ČR project 24-10306S.}

\author{Hans Raj {Tiwary}}{Charles University, Faculty of Mathematics and Physics, Department of Applied Mathematics, Czechia \and \url{https://kam.mff.cuni.cz/~hansraj/}}{hansraj@kam.mff.cuni.cz}{https://orcid.org/0000-0003-1903-1600}{Partially supported by the AGATE project funded  from the Horizon Europe Programme under Grant Agreement No. 101183743.}
\authorrunning{S. Atalig, M. Chrobak, P. Kolman, H. R. Tiwary}

\Copyright{Sunny Atalig, Marek Chrobak, Petr Kolman, Hans Raj Tiwary}

\category{}
\relatedversion{}
\supplement{}

\nolinenumbers
\begin{document}
\maketitle
\begin{abstract}
The spanning tree congestion of a connected graph \(G\) is the minimum, over all spanning trees \(T\) of \(G\), of the maximum number of edges of \(G\) crossing a cut induced by deleting an edge of \(T\). In the spanning tree congestion problem we are asked to compute a spanning tree with minimum congestion.

We describe two sublinear-ratio approximation algorithms 
for the spanning tree congestion problem (STC):
a simple $\Ot(n^{8/9})$-approximation, and a more involved $\Ot(n^{2/3})$-approximation, where  the notation $\Ot$ hides factors polynomial in $\log n$.. 
Until now, no algorithm with sublinear ratio was known. 

We also prove that for any fixed $0<\varepsilon<1$, 
approximation of the spanning tree congestion within $2-\varepsilon$
is NP-hard.
This improves the previous best known lower bound $6/5$.

\end{abstract}

\section{Introduction}\label{sec:intro}
Let $G$ be a connected graph and $T$ a spanning tree of $G$. 
We can imagine routing every edge $uv$ of $G$ along the unique $u$--to--$v$ path in $T$. 
The congestion $c(G,T)$ is the maximum number of these paths using one tree edge, and $\STC(G)$ is the minimum of $c(G,T)$ over all spanning trees $T$ of $G$. 
Equivalently, the congestion of a tree edge $e$ is the number of graph edges in the cut between the two subtrees of $T$ resulting from removing
$e$ from $T$. The tree edge itself is counted.

Spanning tree congestion is difficult to approximate and no sublinear approximation algorithms for general graphs are known. Two useful approaches bound congestion in different ways.
Chandran, Cheung, and Issac~\cite{Chandranetal:18} construct a tree of
congestion $O(\sqrt{mn}\log n)$ in polynomial time, where $n$ and $m$ are the
numbers of vertices and edges. Combining their approach with the recent algorithm 
in~\cite{Hajiaghayietal:26} avoids the $\log n$-factor and yields
a spanning tree of congestion at most $8\sqrt{mn}$
(see Proposition~\ref{prop:cci}). Kolman~\cite{Kolman2026}
gives an $O(\Delta\log^{3/2}n)$-approximation, where $\Delta$ is the maximum
degree, using hereditary bisection. The first bound is useful when the optimum
is large; the second is useful when degrees are small.

In the connected min-max multiway cut problem, we are given a graph $G$ with a set of $t$ terminals $\Gamma \subseteq V(G)$. A connected multiway cut is a partition $V_1,\ldots,V_t$ of $V(G)$ such that for each $i\in[t]$:
i) $G[V_i]$ is connected, and,
ii) $|V_i\cap T|=1$.
The cost of such a cut is the maximum boundary size of any $V_i$. The goal is to find a connected multiway cut of minimum cost. This problem was studied by Kolman and Tiwary \cite{KT:26} who were motivated by its connection to the STC problem.

\paragraph{Our results.}
We first give, in Section~\ref{sec:simple}, a randomized 
$\Ot(n^{8/9})$-approximation for spanning tree congestion;
this is the first algorithm with sublinear approximation ratio.
One of the tools used in this algorithm is an approximation for
the connected min--max multiway cut,  and we provide an $\Ot(\sqrt{t})$-approximation
for it in Appendix~\ref{app:conminmaxcut}, where $t$ is the number of terminals.

Then, in Section~\ref{sec:laminar-alg}, we describe our main result, which is an improvement
of the simple algorithm.
%
%
\begin{theorem}\label{thm:main}
	There is a randomized polynomial-time
	algorithm for spanning tree congestion
	with approximation ratio $\Ot(n^{2/3})$.
\end{theorem}

On the hardness side, in Section~\ref{sec:hardness} we prove that no polynomial-time
$(2-\varepsilon)$-approximation exists for any fixed $0<\varepsilon<1$, unless
$\PP=\NP$. 

\paragraph{Sketch of the algorithms.}
Both approximation algorithms for spanning tree congestion use the following
structure. Identify the high-degree vertices, called \emph{hubs}. Separate them
into connected regions containing one hub each. Within a region, remove its hub
and partition the remaining vertices into connected parts with small boundaries. 
Construct a tree inside each part, join these trees through the hub, and finally join the
regional trees. 
The algorithms differ in the two partitioning phases: how to obtain regions with
only one hub, and how to partition the region with exactly one hub.

Our first algorithm combines two existing algorithms for STC — the spanning-tree
construction of Chandran, Cheung, and Issac, which achieves congestion
\(\Ot(\sqrt{mn})\), and Kolman’s \(\Ot(\Delta)\)-approximation
algorithm — with a classical minimum-cut algorithm and 
our $\Ot(\sqrt{t})$-approximation of the connected min-max multiway partition with $t$ terminals. 
Here, \(n\), \(m\),
and \(\Delta\) denote the number of vertices, the number of edges, and the
maximum vertex degree of the input graph, respectively.

The second algorithm first constructs a new laminar decomposition of the
input graph $G$, taking advantage of the bound on optimal congestion to
control the boundary size of the sets in this decomposition, and the
vertex degrees in each set.
Then it recursively constructs spanning trees of graphs that are quotients
of $G$ with respect to the layers of this decomposition, eventually
producing a low-congestion spanning tree of $G$.

In both algorithms, an important tool is a subroutine that constructs a spanning
tree for a graph with one large degree vertex.
To solve this task, the first algorithm uses an approximation
of min-max connected partition combined with confluent flows.
The second algorithm uses confluent flows as well, but constructs a more
sophisticated connected partition, with additional constraints on the
distribution of edges between different parts. 

The two key ingredients in our $\Ot(n^{2/3})$-approximation, namely
the new laminar decomposition of a graph and the connected partitioning,
might be of independent interest.

\paragraph{Preliminaries.}
For a graph $G$, $V(G)$ (or $V_G$) denotes its vertex set and $E(G)$ (or $E_G$) its edge set.
Throughout the approximation sections, $G$ is a connected simple unweighted
graph, $n=|V_G|\ge2$, $m=|E_G|$, $k=\STC(G)$, and $L=\log n$.
The degree of a vertex $v$ in $G$ is denoted $d_G(v)$.
For disjoint vertex sets $A,B$ in a graph $G$, let $E_G(A,B)$ be the set of
edges between them and $e_G(A,B)=|E_G(A,B)|$. Write $G[A]$ for the subgraph
induced by $A$. An auxiliary graph $F$ may have positive integer edge weights
that we denote by $\omega_F$; for $G$, $\omega_G$ are unit weights. Parallel edges may be combined by adding their weights, and if we contract a subset of vertices we discard loops. Define
\[
\delta_G(A)=E_G(A,V_G\setminus A),\qquad
b_G(A)=\sum_{e\in\delta_G(A)}\omega_G(e).
\]
The weighted
degree is $d_G(v)=b_G(\{v\})$. For an edge set $E'$, write $\omega_G(E')$ for
its total weight, and for disjoint $A,B$ write
$\omega_G(A,B)=\omega_G(E_G(A,B))$.

Let $P_T(u,v)$ denote the unique $u$-to-$v$ path in a spanning tree $T$ of $G$.
Define
\[
c_{G,T}(e)=\sum_{\substack{uv\in E_G\,: \\E(P_T(u,v)) \ni e}}\omega_G(uv),\qquad
\ell_{G,T}(z)=\sum_{\substack{uv\in E_G \,:\\V(P_T(u,v)) \ni z}}\omega_G(uv).
\]
The first quantity is the \emph{edge congestion} of edge $e$ (also called its \emph{load}) in a spanning tree $T$
of a graph $G$; the second quantity is called \emph{vertex load}.
The congestion of a spanning tree $T$ is $c(G,T)=\max_{e\in E(T)}c_{G,T}(e)$. 
A spanning tree of a graph on a single 
vertex has congestion zero. 

We use the following consequence (see Appendix~\ref{app:cci}) 
of the new algorithm in~\cite{Hajiaghayietal:26}, which slightly improves
the construction by Chandran, Cheung and Issac~\cite{Chandranetal:18}.
\begin{lemma}[\cite{Chandranetal:18,Hajiaghayietal:26}]\label{prop:cci}
	For every connected simple unweighted graph $G$ on $n\ge2$ vertices and
	$m$ edges, a spanning tree of congestion at most $8\sqrt{mn}$ can be
	computed in polynomial time.
\end{lemma}

We write $T^*$ for an optimum spanning tree of~$G$.
Every graph edge uses at least one edge of $T^*$, whence
\begin{equation}\label{eq:optimum-density}
	m\le\sum_{e\in E(T^*)}c_{G,T^*}(e)\le k(n-1)\le kn.
\end{equation}
Applying the bound~\eqref{eq:optimum-density} on the spanning tree from 
Proposition~\ref{prop:cci}, we get that its approximation ratio is
$O(\sqrt{mn}/k) = O(n/\sqrt{k})$. Whenever we refer to the \Chandranalg 
tree below, we mean the construction of Proposition~\ref{prop:cci}.

For $b>1/2$, a $b$-balanced cut of $G$ has both sides of size at most $b\cdot|V(G)|$. The
\emph{hereditary bisection} $\operatorname{hb}(G)$ is the maximum, over all
subgraphs $H$ with at least two vertices, of the minimum weight of a
$2/3$-balanced cut of $H$. 

The following lemma roughly says that for a graph
of hereditary bisection~$B$, we can efficiently construct a spanning tree of
congestion $\Ot(B)$.
\begin{lemma}[\cite{Kolman2026}]\label{lem:hereditary}
	Let $H$ be a connected positive-integer-weighted graph on $n$ vertices,
	of total edge weight $M_H$ and hereditary bisection $B$.
	There is an algorithm that in time polynomial in $n+M_H$ constructs
	a spanning tree with edge and 
	vertex loads $O(B\log^{3/2} n)$.
\end{lemma}
The proof (for completeness given in Appendix~\ref{app:proofs}) 	
is a straightforward extension of the analysis of the hereditary-bisection 
based approximation algorithm for spanning tree congestion~\cite{Kolman2026}.

\paragraph{Min-max connected multiway cut.} Given a graph $G=(V,E)$ with positive integer weights and a set
$\Gamma\subseteq V$ of $t$ terminals, partition $V$ into $t$ parts such that
each part is connected and contains exactly one terminal; the objective is to
minimize the maximum weight of  the edges leaving any part of the partition.
The following approximation, proved in Appendix~\ref{app:conminmaxcut},
is used as a tool in the simple algorithm for spanning tree congestion.
\begin{theorem}\label{thm:conminmaxcut}
	There exists a randomized $\Ot(\sqrt t)$-approximation for the Min-max connected
	multiway cut with $t$ terminals. Moreover, the maximum boundary of the constructed
	connected partition is $\Ot(\min\{\sqrt{t}\cdot OPT, \sqrt{M\cdot OPT}\})$ where $M$ 
	is the total edge weight in the graph.
\end{theorem}
An important tool for proving Theorems~\ref{thm:main} and~\ref{thm:conminmaxcut} is confluent
flow~\cite{Chen2007}. We will use the following result.
\begin{theorem}[Theorem 20~\cite{Chen2007}]\label{thm:confluent}
	Given a splittable flow $f$  on a graph with $t$ sinks, there is a strongly 
	polynomial time algorithm that finds a confluent flow satisfying the same demand 
	and with node load at most $(1+\ln t)$-times larger.
\end{theorem}

\section{A simple \texorpdfstring{$\Ot(n^{8/9})$}{O-tilde(n to the power 8/9)}-approximation}
\label{sec:simple}
For a degree threshold $D$, that we will set up later, define
$H=\{v\in V:d_G(v)>D\}$ and $q=|H|$.
Write $H=\{v_1,\ldots,v_q\}$ and call the vertices in $H$ \emph{hubs}.
For now suppose $1\le q$; otherwise Kolman's algorithm~\cite{Kolman2026} gives 
$\Ot(D)$-approximation. 

The algorithm has three main phases: i) separate the hubs into connected regions, 
ii) split each region around its hub, and
iii) construct trees in the resulting smaller parts, attach them to the hub in
their region, and arbitrarily connect the subtrees into a spanning tree of~$G$.

Let $M$ be the size of a multiway cut separating the vertices of $H$;
such a cut of size at most $(q-1)\cdot k$ can be constructed in polynomial
time by performing $q-1$ minimum cuts between pairs of hubs, always within
a current connected region containing both hubs.
(Note that each individual mincut's size is at most $k$.)
Thus, $ M\le(q-1)\cdot k$.
Let $R_1,\ldots,R_q$ be the corresponding connected subgraphs with
$R_i\cap H=\{v_i\}$; then $b_G(R_i)\le M$. We refer to these $R_i$'s as \emph{regions}.

\begin{lemma}
	\label{simple:regions}
	If $D\ge4k$, then $q<2n/D$ (so $q<n$).
\end{lemma}
\begin{proof}
	Since $G$ is simple, each hub $v_i$ has at most $|R_i|-1$ neighbors inside  $R_i$.
	An edge between regions contributes at most two to the sum
	of the hubs' outside degrees. Thus,
	\[
	\sum_i d_G(v_i)\le\sum_i(|R_i|-1)+2M
	\le n-q+2(q-1)k.
	\]
	For $D\ge4k$, we get
	$ qD<\sum_i d_G(v_i)\le n+2qk\le n+qD/2$,
	and hence $q<2n/D$. 
\end{proof}

For a region $R_i$, its \emph{terminals} are the
neighbors of $v_i$ in $R_i$. 
We refer to induced subgraphs of $R_i$ as \emph{sub-regions}. If a sub-region
contains exactly $r$ terminals we call it an \emph{$r$-terminal sub-region}.
We want to partition $R_i\setminus
\{v_i\}$ into connected $1$-terminal sub-regions. Then the spanning trees of these sub-regions can
be joined to $v_i$, each by its unique edge to $v_i$.
\begin{lemma}\label{simple:witness}
	For each $i\in [q]$,
	there is a partition of $R_i\setminus\{v_i\}$ into connected $1$-terminal sub-regions,
	with $b_G(P)\le qk^2$ for every sub-region $P$ in this partition.
\end{lemma}
\begin{proof}
	Fix $i\in [q]$ and write $R=R_i$, $v=v_i$, and note that 
	$b_G(R)\le(q-1)k$. Every connected component of the forest $T^*\setminus\{v\}$ 
	has boundary at most $k$; therefore it contains at most $k$ neighbors
	of $v$ (possibly none). Each connected component of $T^*[R\setminus\{v\}]$ thus
	contains at most $k$ terminals.
	These connected components form our initial partition into sub-regions, that we now gradually modify.
	
	First, for any $r$-terminal sub-region, with $r > 1$, delete from it $r-1$
	edges of $T^*$, thus breaking it into connected $1$-terminal sub-regions.
	It now remains to deal with $0$-terminal sub-regions. We do it as follows.
	While there is a $0$-terminal sub-region in the current partition, find
	one that has an edge connecting it to a $1$-terminal sub-region (this is possible, by the connectivity of $R$)
	and merge it with this $1$-terminal sub-region.
	This will eventually create a partition of $R\setminus\{v\}$ into connected $1$-terminal sub-regions.
	
	By construction, each $1$-terminal sub-region $P$ in this partition has 
	(i) at most $k-1$ edges of $T^\star$ inside $R\setminus \{v\}$ to other partition's sub-regions,
	(ii) at most one edge of $T^\star$  to $v$, and 
	(iii) at most $b_G(R)$ edges of $T^\star$  to $V(G)\setminus R$. As the spanning
	tree congestion of $T^\star$ is $k$, we get $b_G(P)\le k(k-1+1+b_G(R))\le k(k+(q-1)k) = qk^2$.
\end{proof}

Lemma~\ref{simple:witness} bounds the boundaries of an optimal partition of $R_i$ into connected $1$-terminal sub-regions.
To compute an approximate partition,
we can apply Theorem~\ref{thm:conminmaxcut} separately to each connected component of
$G[R_i\setminus v_i]$.	
Each resulting  sub-region $P$ has boundary $\Ot(k\sqrt{mq})$ within $G[R_i\setminus v_i]$, 
has exactly one edge to $v_i$,
and at most $b_G(R_i)$ edges leaving $R_i$. Since
$1+b_G(R_i)\le qk\le k\sqrt{mq}$ for $q<n$ and $m\ge n-1$, we obtain
\begin{equation}\label{simple:piece-boundary}
	b_G(P)=\Ot(k\sqrt{mq}).
\end{equation}

By the following lemma, for a connected subgraph with \emph{bounded}
boundary and \emph{bounded} largest degree, we can construct a spanning
tree with a bounded congestion.
\begin{lemma}\label{simple:internal}
	Let $P\subseteq V(G)$ be nonempty with $G[P]$ connected, and put
	$D_P=\max_{u\in P}d_G(u)$. Then
	Kolman's algorithm~\cite{Kolman2026} constructs a spanning tree $T_P$
	of $G[P]$ with
	\begin{equation}\label{simple:internal-load}
		c(G[P],T_P)=O\!\left(k\max\{D_P,b_G(P)\}\log^{3/2}n\right).
	\end{equation}
\end{lemma}
\begin{proof}
	We start by showing that 
	$\operatorname{hb}(G[P])\le k\max\{D_P,b_G(P)\}$.
	Take any $X\subseteq P$ with $|X|\ge2$. Mark its vertices in $T^*$
	and choose a weighted centroid $z$, so that every component of
	$T^*-z$ contains at most $|X|/2$ marks. Group these components, and
	the singleton $\{z\}$ if marked, into two sides each containing
	between $|X|/3$ and $2|X|/3$ marks. To do so, take a block of size
	at least $|X|/3$ if one exists; otherwise accumulate blocks until
	this threshold is first reached.
	
	Every edge of $G[X]$ crossing this cut has its tree path pass
	through $z$. If $z\in P$, that path uses an edge of $T^*$ incident
	with $z$; there are at most $D_P$ such edges. If $z\notin P$, the
	path uses an edge of $T^*$ leaving $P$; there are at most $b_G(P)$
	such edges. Their loads are at most $k$, giving the claimed balanced
	cut bound. Deleting edges cannot increase the minimum balanced-cut
	cost, so the bound holds for all subgraphs of $G[P]$. The tree
	bound follows from the hereditary-bisection black box,
	Lemma~\ref{lem:hereditary}.
\end{proof}

\subsection{Algorithm and analysis} 
Write \textsc{HereditarySpanTree} for the construction of
Lemma~\ref{simple:internal}, and \textsc{ChandranCheungIssacTree}
for the construction of Proposition~\ref{prop:cci}.

\begin{algorithm}[H]
	\caption{\textsc{SimpleSpanTree}$(G)$}
	\label{alg:simple-span-tree}
	\begin{algorithmic}[1]
		\State $n\gets |V(G)|$, $D\gets 4n^{8/9}$
		\State $H\gets\{v\in V(G):d_G(v)>D\}$, $q\gets |H|$
		\State $T_0\gets\Call{\Chandranalg}{G}$
		\Comment{Proposition~\ref{prop:cci}}
		\If{$q=n$}
		\State \Return $T_0$
		\EndIf
		\If{$q=0$}
		\State $T\gets\Call{HereditarySpanTree}{G}$
		\Else
		\State Write $H=\{v_1,\ldots,v_q\}$
		\State Compute a multiway cut of size $M\le(q-1)k$
		using $q-1$ successive minimum cuts
		\State Let $R_1,\ldots,R_q$ be the resulting connected regions,
		with $R_i\cap H=\{v_i\}$
		\For{$i=1,\ldots,q$}
		\ForAll{connected components $C$ of $G[R_i\setminus\{v_i\}]$}
		\If{$C$ contains at least two terminals}
		\Comment{Terminals are neighbors of $v_i$}
		\State Partition $C$ into connected $1$-terminal
		sub-regions using Theorem~\ref{thm:conminmaxcut}
		\EndIf
		\EndFor
		\ForAll{sub-regions $P$ obtained for $R_i$}
		\State $T'_P\gets\Call{HereditarySpanTree}{G[P]}$
		\EndFor
		\State $T_i\gets$  tree obtained by joining each $T'_P$
		to $v_i$ through its unique terminal
		\EndFor
		\State $T\gets$ spanning tree of $G$ obtained by joining
		$T_1,\ldots,T_q$ using multiway-cut edges
		\EndIf
		\State \Return a tree of minimum congestion among $T$ and $T_0$
	\end{algorithmic}
\end{algorithm}

\begin{figure}[tb]
	\centering
	\includegraphics[width=0.7\linewidth]{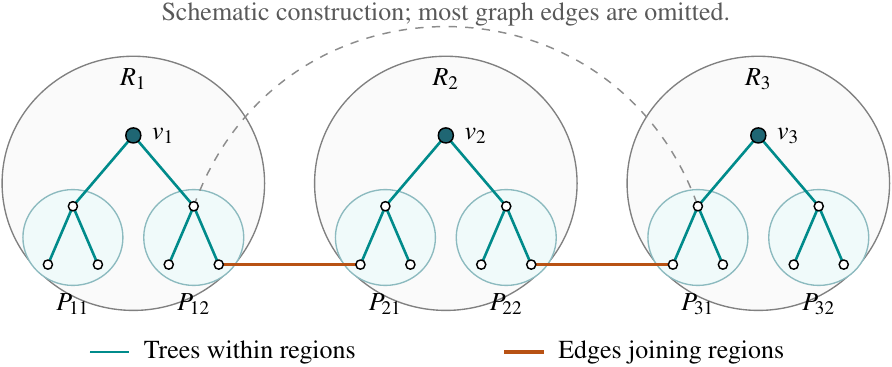}
	\caption{The simpler algorithm. Each region $R_i$ contains one hub $v_i$.
		The connected parts $P_{ij}$ each contain one neighbor of $v_i$; their internal trees are joined to $v_i$ through these neighbors. 
		A tree of the region quotient then joins the regional trees. Solid edges show the constructed tree; 
		the dashed edge represents another graph edge. Most graph edges are omitted.}
	\label{fig:simple}
\end{figure}

Figure~\ref{fig:simple} illustrates the construction. 

\begin{theorem}\label{simple:approx}
	The approximation ratio of the above described randomized algorithm is 
	$\Ot(n^{8/9})$.
\end{theorem}
\begin{proof}
	Suppose first that $1\le q<n$. Every connected $1$-terminal sub-region $P$ constructed 
	in step~(2) has largest degree
	$D_P\le D$. When the internal tree of $P$ is joined to the hub, edges
	with an endpoint outside $P$ add at most $b_G(P)$ to the congestion of its edges.
	The edge joining $P$ to the hub has load at most $b_G(P)$, which is at most $\Ot(k\sqrt{mq})$ by \eqref{simple:piece-boundary}.
	When the regional trees are joined, only the $M$ graph edges
	between regions can add load; a joining tree edge has load at most
	$M$. Thus, by the above bounds and
	\eqref{simple:internal-load}, the output spanning tree $T$ has congestion at most
	\[
	k\,\Ot(D+k\sqrt{mq})+M.
	\]
	For $D\ge4k$, Lemma~\ref{simple:regions} gives $q<2n/D$.
	Also $m\le k(n-1)$, by~(\ref{eq:optimum-density}).
	The approximation ratio is therefore
	\begin{equation}\label{simple:ratio}
		\Ot\!\left(D+ {nk^{3/2}}/{\sqrt D}\right).
	\end{equation}
	Here $M\le(q-1)k$ is absorbed by the displayed bound.
	The Chandran--Cheung--Issac construction of Proposition~\ref{prop:cci}
	has approximation ratio $O(n/\sqrt k)$.
	
	If $k\le n^{2/9}$, then $D\ge4k$, and both terms
	in~\eqref{simple:ratio} are $\Ot(n^{8/9})$.
	If $k>n^{2/9}$, the Chandran--Cheung--Issac ratio is
	$\Ot(n^{8/9})$. The case $q=0$ has ratio
	$O(D\log^{3/2}n)$ by Lemma~\ref{simple:internal}. If $q=n$, 
	the congestion of any edge adjacent to a leaf is at least
	$d_{G}(u)>D$, so $k\ge d_{G}(u)>D$, so again the Chandran--Cheung--Issac bound suffices. 
\end{proof}

The preceding proof identifies two distinct losses of the simple algorithm.

\emph{Separating the hubs.} The first partition has total cut
size at most $(q-1)k$, and even some individual region may have boundary of
that order. This leads to the $qk^2$ bound in
Lemma~\ref{simple:witness}. In contrast, our improved algorithm will find
small-boundary regions progressively in several rounds: in each round,
it constructs regions with boundary $O(k)$ around at least half of the 
hubs, then contracts these regions into single vertices, and repeats
the same step on the modified graph. 
This way, after at most logarithmically many rounds, it separates all hubs.

\emph{Controlling the smaller parts.} A connected min--max
multiway cut controls $b_G(P)$ but gives no further guarantee about the
edges inside $P$. The internal-tree bound therefore pays
$k\max\{D,b_G(P)\}$. The main algorithm instead constructs a
partition $\mathcal P$ that also controls, for every small-boundary
set $C$, the edges of $G[C]$ between different parts of $\mathcal P$.
This simultaneous guarantee is what permits efficient routing
between parts while Kolman's algorithm constructs their internal
trees.

\section{An $\Ot(n^{2/3})$ Approximation Via a Laminar Decomposition}\label{sec:laminar-alg}
We first describe the construction for a degree threshold $D\ge4k$; 
the algorithm will try powers of two, without knowing $k$. 
It tries $D=2^i$ for $2\le i\le\lceil\log_2(8m)\rceil$, discards
failed guesses, and retains the least-congested tree obtained.

For disjoint vertex sets $A,B$ in $G$, let
$\mincut_G(A,B)=\min\{b_G(X):A\subseteq X\subseteq V_G\setminus B\}$.

As in Section~2, put $H=\{v:d_G(v)>D\}$ and call its vertices \emph{hubs}. 
The algorithm has two phases. First, it constructs a laminar decomposition of $G$
which is a laminar
family $\CL$ of connected sets that separate the hubs; each set in
the family has boundary $O(D)$. Next, the algorithm
processes the laminar tree of $\CL$ and gradually constructs a 
spanning tree.

\paragraph{\textbf{Graph Quotients.}}
Given a graph $G$, 
a \emph{bag} is a connected set of its vertices. For a partition
$\calP$ of $V_G$, the \emph{quotient} of $G$ by $\calP$ is the graph 
$F=G/{\calP}$ that has vertex set $\calP$ and where distinct sets $A, B\in \calP$ 
are adjacent if and only if some vertex in $A$ is adjacent in $G$ to some vertex in $B$;
the edge $AB$ has weight $\omega_F(AB)=e_G(A,B)$.
For a weighted input graph, this weight is instead the sum of the
weights of its edges between $A$ and $B$; this convention applies
when contracting a quotient again.

For a collection $\CS$ of disjoint sets of $G$, we write $G/{\CS}$
for $G/{\CS'}$ where 
$\CS'=\CS\cup\{\{v\}:v\in V_G\setminus\bigcup_{S\in\CS}S\}$,
that is, we view the disjoint collection $\CS$ as a representation of the
partition of $V_G$ that contains the sets in $\CS$ and in which all vertices
not contained in any bag in $\CS$ appear as size-one sets.
For notational simplicity, in quotient graphs, we often identify each 
singleton bag $\{v\}$ with $v$.
We say that $G/{\CS}$ is a quotient of $G$ \emph{at level $1$}, and if $F$ is
a quotient of $G$ at level $i\geq 1$ and $\CS'$ is a collection of disjoint sets of
$F$, then $F/\CS'$ is a quotient of $G$ \emph{at level $i+1$}.

One can naturally extend the notion of a quotient for a laminar family $\CL$ as follows. Recall that a family $\CL$ is \emph{laminar} if any two of its members are
disjoint or one contains the other; the depth of $\CL$ is the maximum number
of members in an inclusion chain. 
We denote the set of inclusion-wise maximal members of $\CL$ by $\CL_1$ and -- if it is unique -- call it the \emph{root} of $\CL$. By $\CL_2$ we denote the members that are inclusion-wise maximal in $\CL\setminus \CL_1$. If $\CL_1$ is the root then $\CL_2$ are the maximal sets in $\CL$ that are properly contained in $\CL_1$. We will view $\CL_2$ as a collection of disjoint subsets -- either of $\CL_1$ or the entire vertex set depending on the context -- and denote by $G/\CL$ the \emph{quotient} $G/\CL_2$.

\subsection{Construction of the Laminar Graph Decomposition}
Let $G$ be a graph. For $S\subseteq V_F$, where $F$ is a quotient of $G$, let
$S^\uparrow=\bigcup_{A\in S}A$ denote the union of the sets corresponding
to the vertices in $S$. For $S\subseteq V_F$, if $F$ is a quotient of $G$ at level $1$,
then let $S^*=S^\uparrow$, and if $F$ is a quotient of $G$ at level $i\geq 2$,
then let $S^*=\bigcup_{A\in S} A^*$. 
For a collection $\CC$ of subsets of $V_F$,
define $\mathcal C^\uparrow=\{S^\uparrow:S\in\CC\}$. Informally speaking, we will construct a hierarchy of quotient graphs and for $S\subseteq V_F$ for some quotient $F$ in this hierarchy every vertex $v\in V_F$ represents a subset of vertices in the parent graph that were contracted. $S^\uparrow$ then are the vertices that are one level up and that were contracted to get $S$ while $S^*$ are the vertices in the original graph. For example, suppose we start with $G$. Let $F$ be a quotient of $G$ and let $H$ be a quotient of $F$. Then, for $S\subseteq V_H$, $S^\uparrow$ are vertices in $V_F$ that were contracted to get $S$ whereas $S^*$ are the vertices in $V_G$ that were eventually contracted to get $S$.

Given a weighted graph $G$ and a subset $H\subset V_G$ of size at
least $2$, the \emph{minimum isolating cuts} for $H$ is a collection of disjoint sets
$\{S_h:h\in H\}$ satisfying for each $h\in H$, $S_h\cap H=\{h\}$ and 
$b_G(S_h)=\mincut_G(\{h\},H\setminus\{h\})$.
Minimum isolating cuts for $H$, 
in which each set $S_h$ induces a connected subgraph, can be constructed in
polynomial time by repeated calls to a max-flow algorithm (e.g.,~\cite{LP:21});
in the following, we refer to this algorithm as the \isolalg. Its input
is the pair $(G,H)$, and the output is a collection of 
the minimum isolating cuts for $H$.

Throughout the laminar construction, contracting an isolating set
$S_h$ retains the name $h$ for the resulting vertex; all other
vertices keep their names. We use the same names in the local
graphs $G[R]/\CL$ that are used in the construction. 
Their vertex sets are therefore subsets of~$V_G$.

In our laminar family, the boundary of each non-root member set 
will be a minimum cut in $G$ that separates (at the time when this set is being added to the family)
one yet non-contracted hub from all other yet non-contracted hubs. 
The boundary of the root $V_G$ is empty.
The following lemma follows directly from sub-modularity of cuts (cf.
Lemma 3.1 in the paper~\cite{LP:21});
for the sake of completeness, we give its proof (in Appendix).
\begin{lemma}\label{lem:mincutpreservation}
	Let $H\subseteq V$, $H'\subseteq H$, and let
	$\mathcal S=\{S_h:h\in H'\}$ be pairwise disjoint sets satisfying,
	for each $h\in H'$, $ S_h\cap H=\{h\}$ and
	$b_G(S_h)=\mincut_G(\{h\},H\setminus\{h\})$.
	Then, for every pair of disjoint sets $A,B\subseteq H\setminus H'$,
	\begin{equation}\label{eq:remaining}
		\mincut_{G/{\mathcal S}}(A,B)=\mincut_G(A,B).
	\end{equation}
	More generally, if we identify the contracted vertex $S_h$ with the corresponding vertex 
	$h$ and set $R=(V\setminus\bigcup_{h\in H'}S_h)\cup H'$,
	then~\eqref{eq:remaining} holds for all disjoint $A,B\subseteq R$.
\end{lemma}
\begin{proof}
	Identify each contracted vertex $S_h$ with $h$ and put
	$ R=\Bigl(V\setminus\bigcup_{h\in H'}S_h\Bigr)\cup H'$.
	We prove a slightly more general statement, namely that 
	equality~\eqref{eq:remaining} holds for all disjoint $A,B\subseteq R$.
	
	For disjoint sets $A,B\subseteq R$, fix a minimum $A$--$B$ cut $X$ in $G$, that is,
	$A\subseteq X\subseteq V\setminus B$ minimizing $b_G(X)$.
	We are going to construct an $A$--$B$ cut in $G/{\mathcal S}$ of the same size.
	Process the sets $S_h$ from $\mathcal S$ one at a time. If $h\in X$, replace
	$X$ by $X\cup S_h$; note that the set $X\cap S_h$ separates $h$ from
	$H\setminus\{h\}$. Thus, by the optimality of $S_h$, and by sub-modularity of 
	cuts we get
	\[
	b_G(X\cup S_h)+b_G(X\cap S_h)
	\le b_G(X)+b_G(S_h)
	\le b_G(X)+b_G(X\cap S_h).
	\]
	Hence this replacement does not increase the boundary. If $h\notin X$,
	apply the same argument to $V\setminus X$ and use symmetry of the
	cut function: replacing $X$ by $X\setminus S_h$ also does not
	increase its boundary.
	
	Each replacement puts the entire set $S_h$ on the side of $X$ containing
	$h$. Since $S_h\cap R=\{h\}$, no terminal of $A$ or $B$ changes
	sides, and $X$ remains an $A$--$B$ cut. 
	Disjointness of the sets $S_h$ ensures that subsequent replacements do not
	split a set already processed. For the resulting cut $X$, $X/{\mathcal S}$ 
	is therefore a
	cut of $G/{\mathcal S}$ of weight at most $\mincut_G(A,B)$.
	Conversely, every cut of the quotient induces a cut of $G$ with
	exactly the same weight and the same terminal separation. 
\end{proof}

\begin{algorithm}[H]
	\caption{{\lamalg}$(F,D)$}
	\label{alg:laminar-construction}
	\begin{algorithmic}[1]
		\State $H_F\gets\{v\in V_F:d_F(v)>D\}$
		\If{$|H_F|\le 1$}
		\State \Return $\{V_F\}$
		\Comment{backtracking will convert $\{V_F\}$ to $V_G$}
		\EndIf
		\State $\CC\gets\Call{\isolalg}{F,H_F}$
		\State $\CC_1\gets\{S\in \CC : b_F(S)\le D\}$
		\If{$\CC_1=\emptyset$}
		\State \Return \textsc{Fail} \Comment{unsuccessful guess of $D$}
		\EndIf
		\State $F_1\gets F/{\mathcal C_1}$
		\State $\CC_2\gets\Call{\lamalg}{F_1,D}$
		\If{$\CC_2=\textsc{Fail}$}
		\State \Return \textsc{Fail}
		\EndIf
		\State \Return $\CC_1\cup \CC_2^\uparrow$
	\end{algorithmic}
\end{algorithm}

\begin{lemma}[Laminar Graph Decomposition]\label{lem:family}
	For every $D\ge4k$, the algorithm \lamalg{} applied on the graph $G$
	constructs a laminar family $\CL$ with the following properties:
	\begin{enumerate}
		\item 
		The depth of $\CL$ is $O(\log(|H|+2))$,
		where $H = \{v\in V : d_G(v)>D\}$.
		\item For each $S\in \CL$, 
		a) $G[S]$ is connected, and, 
		b) $b_G(S)\le D$.
		\item For each $S\in \CL$,
		$ \left|\left(S\setminus\bigcup_{A\in\CL,A \subsetneq S}A\right)\cap H\right| \le 1$.
		\item $V_G\in \CL$.
	\end{enumerate}
\end{lemma}

\begin{proof}
	Each contracted set is connected and has boundary at most $D$,
	so the weighted degree of the new vertex corresponding to it is at most $D$. 
	Contractions of
	other bags preserve this degree. Original singleton vertices also
	retain their original degrees. Hence, $H_F$ consists exactly of
	the original hubs not yet contracted, and $H_F\subseteq H$.
	
	Consider a recursive call \textsc{LaminarConstruction}$(F,D)$, and let
	$q=|H_F|\ge2$. Any two hubs in $H_F$ can be separated in $G$
	by a fundamental cut of an optimum spanning tree, of cost at most $k$.
	More strongly, by deleting $q-1$ edges in the optimum tree, we can separate the $q$ hubs into
	disjoint sets $\{S_h^*\}_{h \in H_F}$ where $\sum_{h\in H_F} b_G (S_h^*) \le 2(q-1)k$.
	Lemma~\ref{lem:mincutpreservation} preserves these cut values in $F$. 
	Consequently,
	$ \sum_{h\in H_F}b_F(S_h)\le2(q-1)k\le qD/2$, implying that
	fewer than $q/2$ isolating sets for $H_F$ have boundary greater than $D$.
	It follows that for at least half of the hubs from $H_F$,
	the connected subsets $S_h$ get contracted, and the number of hubs
	decreases by a factor of two with each recursion level,
	implying that the depth of $\CL$ is $O(\log(|H|+2))$. This proves
	Property~1.
	
	Sets contracted in one round are disjoint, and a later contracted bag
	either contains any earlier bag or is disjoint with it. This proves laminarity. 
	Connectedness is preserved under lifting connected sets of connected
	bags. The identity $b_G(S^*)=b_F(S)$ 
	gives Property~2b 
	for contracted sets, and $b_G(V_G)=0$ gives it for the root.
	
	By construction, at every level of recursion, every set $S\in \CC$ from the
	collection of minimum isolation cuts for $H_F$ in $F$,
	contains exactly one hub from $H_F$.  This implies Property~3.
	
	The single-set family $\{V_F\}$ returned in Line~3 will be
	gradually expanded (in line 12) in the backtracking phase of the recursion
	and eventually converted to set $V_G$, proving Property~4.
\end{proof}

\subsection{Construction of the Spanning Tree}

Let $\CL$ be the laminar family constructed by \lamalg$(G,D)$, and for $S\in \CL$,
let $\CL(S)$ be the set of \emph{all} subsets of $S$ that belong to $\CL$, including $S$. 
We assume that two algorithms are available: \textsc{HereditarySpanTree},  
the approximation algorithm based on hereditary bisection~\cite{Kolman2026}
(see Lemma~\ref{lem:hereditary}),
and \textsc{UniqueHighDegreeSpanTree}, the algorithm of
Lemma~\ref{lem:local}. 
The latter receives the graph $F$, the original high-degree
vertex $h$, and the parameter $D$;
the hereditary routine receives the graph $F$ with all degrees 
at most~$D$. 
By $G[\CL]$ we denote the subgraph of $G$ induced by all the elements in the sets of
the laminar family  $\CL$.

\begin{figure}[htb]
	\centering
	\includegraphics[width=\linewidth]{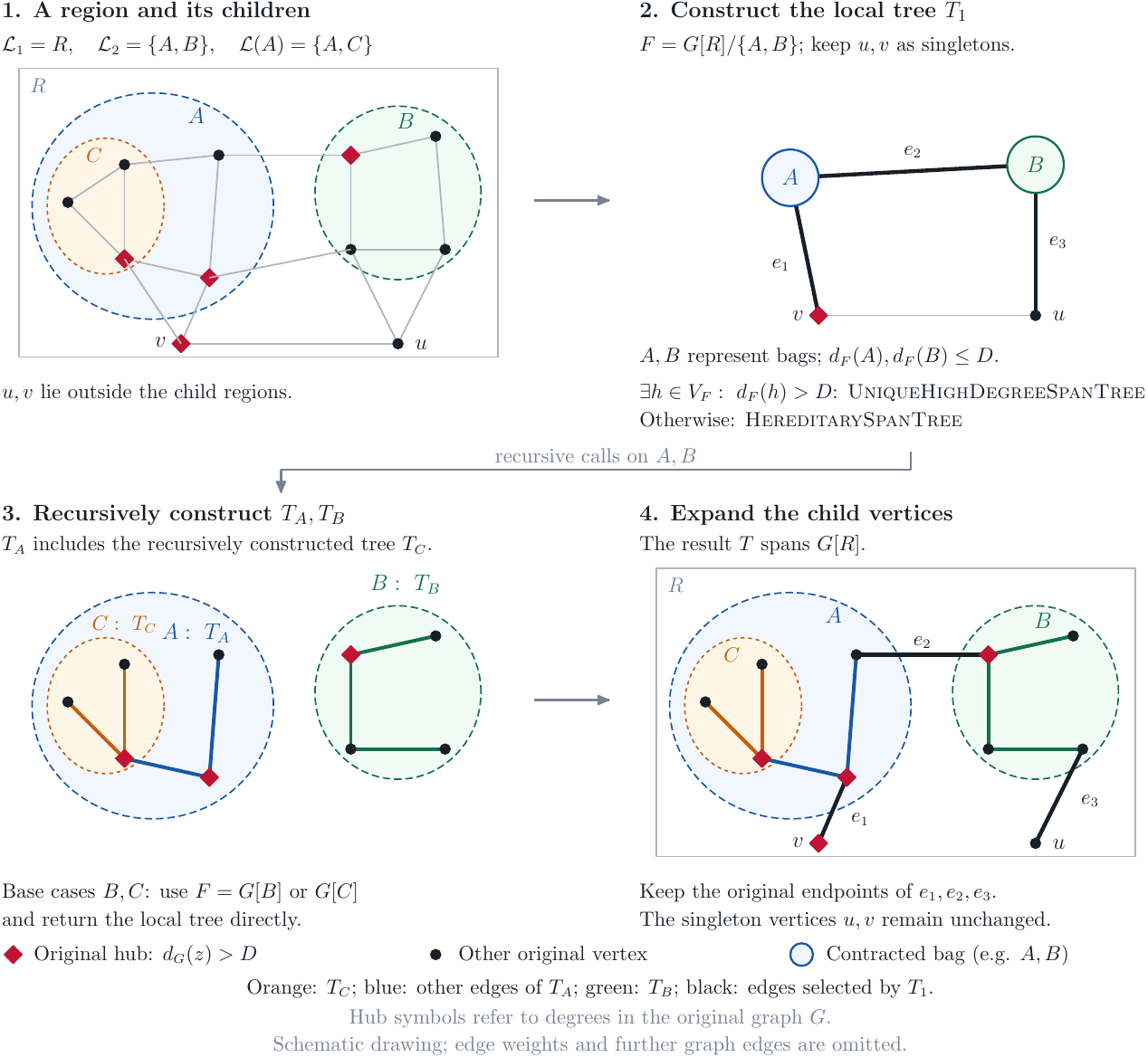}
	\caption{Algorithm~\ref{alg:laminar-span-tree} on a region $R$ with
		children $A,B$ and a further child $C\subset A$. Construct the spanning
		tree $T_1$ for $F=G[R]/\{A,B\}$, recursively obtain the spanning trees for
		$A$ and $B$,
		and then replace the vertices $A$ and $B$ in $T_1$ by their spanning trees.
		The singletons $u,v$ remain unchanged. Red diamonds mark
		hubs of the original graph $G$; the large circles in the quotient
		represent contracted regions.}
	\label{fig:algorithm2}
\end{figure}

\begin{algorithm}[H]
	\caption{\textsc{LaminarSpanTree}$(\CL,D)$}
	\label{alg:laminar-span-tree}
	\begin{algorithmic}[1]
		\If{$\operatorname{depth}(\CL)=1$}
		\State $F\gets G[\CL]$
		\Else
		\State $F\gets G[\CL]/\CL$
		\EndIf
		\If{there is $h\in V_F$ with $d_F(h)>D$}
		\State $T_1\gets\Call{UniqueHighDegreeSpanTree}{F,h,D}$
		\Else
		\State $T_1\gets\Call{HereditarySpanTree}{F}$
		\EndIf
		\If{$\operatorname{depth}(\CL)=1$}
		\State \Return $T_1$
		\EndIf
		\ForAll{$S\in\CL_2$}
		\State $T_S\gets\Call{LaminarSpanTree}{\CL(S),D}$
		\EndFor
		\State $T\gets T_1$ with each vertex corresponding to $S\in\CL_2$ replaced by $T_S$, and each edge $AB$ in $T_1$ replaced by an edge from $E_{G[\CL]}(A,B)$
		\label{alg:combination}
		\State \Return $T$
	\end{algorithmic}
\end{algorithm}
In step~\ref{alg:combination}, an arbitrary edge from $E_{G[\CL]}(A,B)$ can be used. Figure \ref{fig:algorithm2} illustrates the algorithm.

The next lemma bounds the congestion of the spanning tree constructed in
step~\ref{alg:combination} by combination of $T_1$ with the trees $T_S$.
\begin{lemma}\label{lem:expansion}
	Let $Q$ be a quotient of $G$ and $T$ a spanning tree of $Q$.
	Let $a\in V_Q$ have bag $A=\{a\}^*$.
	Refine $A$ into a connected partition $\calP_A$, and
	let $Q'$ be the refined quotient of $G$.
	Choose a spanning tree $T_A$ of $J_A=G[A]/\calP_A$.
	Replace $a$ in $T$ by $T_A$, attaching every old incident tree edge
	to the cell containing the endpoint of its selected original edge.
	The resulting graph $T'$ is a spanning tree of $Q'$. For every new
	tree edge $e$ inside $A$ and every vertex $b\in V_{J_A}$,
	\begin{align}
		c_{Q',T'}(e)&\le c_{J_A,T_A}(e)+\ell_{Q,T}(a),\label{eq:expand-edge}\\
		\ell_{Q',T'}(b)&\le\ell_{J_A,T_A}(b)+\ell_{Q,T}(a).
		\label{eq:expand-vertex}
	\end{align}
	Surviving old tree edges and other old vertices retain their loads.
	The assertions also hold for simultaneous expansion of disjoint bags,
	and for any connected positive-weight graph in place of~$G$.
\end{lemma}
\begin{proof}
	By construction, $T'$ is a spanning tree. 
	Newly visible graph edges inside $A$ have both
	endpoints there; their routes stay in $T_A$ and give its local
	loads. Any previously visible graph edge whose new route uses an
	edge or vertex inside $A$ had a quotient route visiting $a$.
	Their total weight is therefore at most $\ell_{Q,T}(a)$.
	Projecting routes onto $T$ also proves that other old loads do not
	change. The same argument applies to simultaneous expansions.
\end{proof}

The following lemma is a slight generalization of Lemma~\ref{simple:internal};
its proof is in the Appendix.
\begin{lemma}[Internal trees]\label{lem:internal}
	Let $F$ be the local graph at a call of Algorithm~2 with $D\ge4k$.
	For every nonempty $C\subseteq V_F$ such that $F[C]$ is connected
	and $d_F(v)\le D$ for all $v\in C$, \textsc{HereditarySpanTree}$(F)$
	constructs a spanning tree of $F[C]$
	with edge and vertex loads, measured in $F[C]$, at most
	\begin{equation}\label{lem:internal-bound}
		O\!\left(k\bigl(D+b_F(C)\bigr)L^{3/2}\right).
	\end{equation}
	Thus, if in addition $b_F(C)=O(D)$, then 
	\textsc{HereditarySpanTree}$(C)$ returns a spanning tree with
	edge and vertex loads $O(kDL^{3/2})$.
\end{lemma}

\subsection{Spanning Trees for Subgraphs with one High Degree Vertex}\label{sec:onehigh}
In this subsection,  $F$ is a connected positive-integer-weighted
graph obtained as a quotient of a subgraph of $G$, 
and $U\subsetneq V_F$ is a non-empty vertex set. For a partition
$\calP$ of $U$ and $C\subseteq U$, define
\[
x_{\calP}(C)=\omega_F\bigl(\{uv\in E_F:u,v\in C
\text{ belong to different parts of }\calP\}\bigr) \ ;
\]
$x_{\calP}(C)$ measures how much the set $C$ intersects with the
boundaries of the partition~$\CP$.
In this subsection, 
we first describe how to obtain a ``nice'' partition of a subset of vertices,
and then, how to use the partition to construct a spanning tree of $F[U]$ with bounded congestion.

\begin{theorem}[Partition with controlled boundaries]\label{thm:cancellation}
	Let $N=|V_F|$ be the number of vertices in $F$, $M=\sum_{e\in E_F}\omega_F(e)$ 
	be the total edge weight of $F$, $\beta, b_0, d$ and $B_0$ parameters 
	satisfying $\beta=\gamma\log^{3/2}(N+M)$, for a sufficiently large constant $\gamma>0$ 
	and $B_0=\max\{d,\lceil\beta b_0\rceil\}$.
	If every vertex of $U$ has degree at most $d$,
	then one can construct, by a randomized polynomial-time algorithm, 
	a partition $\calP$ of $U$ into connected sets such that
	%
	\begin{align}
		b_F(A)&\le2B_0 && \forall A\in\CP,\label{eq:cancel-part-boundary}\\
		x_{\calP}(C)&\le4\beta b_F(C)
		&& \forall C\subseteq U,\ b_F(C)\le b_0 \ .
		\label{lem:cancel-cancellation}
	\end{align}
\end{theorem}

In the proof of Theorem~\ref{thm:cancellation}, we will need the following
algorithm following from the bisection results of Feige and Krauthgamer~\cite{FeigeKrauthgamer2006} (Subsections 5.2 and 5.4);
for completeness, we describe the reduction in Appendix~\ref{app:quota}.	

\begin{proposition}[Heavy set with small boundary~\cite{FeigeKrauthgamer2006}]\label{prop:quota}
	Let $Q$ be a connected undirected graph on $N$ vertices, with
	nonnegative integer edge capacities and a distinguished vertex $o$.
	Let $\nu$ be a nonnegative polynomially bounded integer vertex weight with $\nu(o)=0$; set
	$W=\nu(V_Q)$, $\beta=O(\log^{3/2}(N+W))$, and let $1\le\ell\le W$ be an integer.	
	There is a randomized polynomial-time algorithm which 
	finds a set $A\subseteq V_Q\setminus\{o\}$ satisfying
	\begin{equation*} 	
		\nu(A)\ge\ell,\qquad
		b_Q(A)\le\beta\min\{b_Q(C):o\notin C,\ \nu(C)\ge\ell\} \ .
		\qquad 
	\end{equation*}
\end{proposition}
Note that the set $A$ may be disconnected. 

\begin{proof}[Proof of Theorem~\ref{thm:cancellation}]
	For a partition $\CP$ of $U$, we define its \emph{potential} as
	\begin{equation}\label{lem:cancel-potential}
		\Phi(\calP)=\sum_{A\in\calP}b_F(A)
		=b_F(U)+2x_{\cal P}(U)
	\end{equation}
	We start the construction with $\CP=\CP_0=\{\{u\}:u\in U\}$ (i.e., each vertex has its own part in $\CP$).
	Note that at this point, for each $A\in \CP$, $b_F(A)\leq 2B_0$; this a property, 
	called the \emph{boundary invariant}, that we will maintain.
	We aim at decreasing the potential as much as possible, and we will do so by a sequence of 
	local changes, called swaps. 
	For $Z\subseteq U$, a $Z$-\emph{swap of} $\CP$ is an operation that changes $\CP$ into a new partition 
	$\CP'$ as follows: For each $A\in\CP$ that has a non-empty intersection with $Z$, 
	$A$ is removed from $\CP$, and all connected components of $F[A\setminus Z]$ are added to $\CP$, 
	and also all connected components of $F[Z]$ are added to $\CP$. 
	Observe that only edges that i) were previously inside a part $A\in \CP$, and ii) belong to $\delta_F(Z)$,
	become edges between parts of $\CP'$, 
	while all edges inside $Z$ that cross different parts of $\CP$ will not cross any boundary of $\CP'$.
	Hence,
	\begin{equation}\label{lem:cancel-swap}
		\begin{split}
		\Delta\Phi = \Phi(\CP')-\Phi(\CP)
		&=2\bigl(\omega_F(\text{edges inside any $A\in \CP$ crossing }Z)-x_{\calP}(Z)\bigr)\\
		&\le2\bigl(b_F(Z)-x_{\calP}(Z)\bigr).
		\end{split}
	\end{equation}
	
	A \emph{plain swap} is a $Z$-swap performed on a set $Z$ satisfying
	\begin{equation}\label{eq:cancel-accept}
		b_F(Z)\le B_0,\qquad x_{\calP}(Z)>b_F(Z).
	\end{equation}
	
	After a swap operation, the boundary invariant might be violated: there might be $A\in \CP'$
	with $b_F(A)>2B_0$.  To
	make it possible to fix it, we maintain for every $A \in\CP'$ a connected set $Q_A$,
	called an \emph{enclosing set of} $A$, with the property
	\begin{equation}\label{lem:cancel-defining}
		A\subseteq Q_A\subseteq U,\qquad b_F(Q_A)\le B_0.
	\end{equation}
	At the very beginning (i.e., for the trivial partition into singletons) we set, for each
	$A\in \CP$, $Q_A=A$. After performing a $Z$-swap, we define the enclosing sets for the new sets 
	in the partition as follows: for each connected component of $A\setminus Z$, considered earlier,
	we define its enclosing set as $Q_A$ (i.e., it inherits the enclosing set of $A$), and 
	for each connected component $C$ of $Z$, we set $Q_C=C$.
	
	A \emph{repair swap} is a $Q_A$-swap for a set $A$ with $b_F(A)>2B_0$.
	Since $A$ is a single part of $\mathcal{P}$ and $Q_A \setminus A$ is disjoint from $A$, 
	the edges between $A$ and $Q_A\setminus A$ 
	join different parts in the partition before performing this swap,
	\begin{equation}\label{eq:aux}
		x_{\calP}(Q_A)\ge \omega_F(A,Q_A\setminus A)
		\ge b_F(A)-b_F(Q_A).
	\end{equation}
	Thus, performing the $Q_A$-swap changes 
	the potential (due to~\ref{lem:cancel-swap} and~\ref{eq:aux}) by
	\begin{equation*}
		\Delta\Phi\le 2(b_F(Q_A)-x_\CP(Q_A)) 
		\leq 2 [ b_F(Q_A)-(b_F(A)-b_F(Q_A)) ]
		=  4b_F(Q_A)-2b_F(A)<0 .
	\end{equation*}
	Both plain and repair swaps decrease the nonnegative integer
	potential by at least two. Initially, it was at most $2M$, so there will
	be at most $M$ operations. 
	
	To complete the description of the sequence of swaps that we use, a few 
	more terms are needed.
	For each $v\in U$, let $A(v)$ be its part in the current partition, and define
	\[
	\nu(v)=\sum_{\substack{vw\in E_F\\w\in U,\ A(w)\ne A(v)}}
	\omega_F(vw) ;
	\]
	$\nu(v)$ is the weight of edges from $v$ to other parts of the current partition of $U$.
	For $C\subseteq U$, put $\nu(C)=\sum_{v\in C}\nu(v)$. Then
	\begin{equation}\label{eq:cancel-mark-identity}
		\nu(C)=2x_{\calP}(C)+\nu_{\partial}(C),
	\end{equation}
	where $\nu_{\partial}(C)$ is the total weight of edges between $C$ and $U\setminus C$ 
	that connect different parts of the current partition; note that for each $C\subseteq U$,
	$\nu_{\partial}(C)\le b_F(C)$.
	
	Let $F_o=F/\{V_F\setminus U\}$, and let $o$ be the vertex obtained by the contraction
	of $V_F\setminus U$. Note that $\forall A\subseteq U$,
	$       b_{F_o}(A)=b_F(A)$.

	Here is the complete algorithm constructing the desired partition. Start with $\CP=\CP_0$, and
	then repeat:
	\begin{enumerate}
		\item While there is $A\in \CP$ with $b_F(A) > 2B_0$, perform $Q_A$-swap.
		\item Compute the weight function $\nu$. If $\nu(U)=0$, STOP.
		\item Set $\nu(o)=0$. For every $q$ equal to a power of two, $q\le\nu(U)$, apply
		Proposition~\ref{prop:quota} on $F_o$ with the weight function $\nu$ and $\ell=q$.
		\item Swap in any returned set satisfying~\eqref{eq:cancel-accept} and
		repeat. If none satisfies it, STOP.
	\end{enumerate}
	
	The repair-swaps ensure the property~\eqref{eq:cancel-part-boundary}. 
	For a contradiction assume that the property~\eqref{lem:cancel-cancellation}
	is not satisfied when the algorithm stops, that is, 
	some $C\subseteq U$ satisfies $b_F(C)\le b_0$ and
	$x_{\calP}(C)>4\beta b_F(C)$. The strict positivity of $x_{\calP}(C)$ implies,
	by~\eqref{eq:cancel-mark-identity}, that $\nu(C)>0$. There is a value $q$
	that is a power of two, was tested by the algorithm, and 
	satisfies $\nu(C)/2<q\le\nu(C)$. 
	Thus, the algorithm of Proposition~\ref{prop:quota} finds a 
	set $Z$ satisfying $b_F(Z)\le\beta b_F(C)\le B_0$
	and
	\[
	\nu(Z)\ge q>\frac{\nu(C)}2\overset{\eqref{eq:cancel-mark-identity}}{\ge} x_{\calP}(C)
	>4\beta b_F(C)\ge4b_F(Z).
	\]
	Combining $\nu(Z) > 4b_F(Z)$ with $\nu_{\partial}(C)\le b_F(C)$ and
	equation~\eqref{eq:cancel-mark-identity} gives
	$x_{\calP}(Z)>\tfrac32b_F(Z)$, contradicting that no set was found in step 4
	(note that if $\nu(U)=0$, no edges join different parts inside $U$, and
	the conclusion is immediate).
	
	There are at most $M$ swaps, one final unsuccessful
	run of the algorithm from Proposition~\ref{prop:quota}, 
	and $O(\log(2M))$ calls of Proposition~\ref{prop:quota} per search. 
	All weights and the parameter $\ell$ are of size at most $2M$. 
	Amplify each call by $O(\log(N+M))$ independent repetitions
	and take a union bound over these
	polynomially many calls, conditional on the preceding choices.
	With high probability every call has the claimed approximation factor,
	so the stopping argument applies simultaneously to all sets $C$.
	Since $d\ge1$, the parameters give
	$B_0=O(d+b_0\log^{3/2}(N+M))$ and
	$4\beta=O(\log^{3/2}(N+M))$.
\end{proof}

\begin{lemma}[Constructing a spanning tree from Theorem~\ref{thm:cancellation}]\label{lem:local}
	Let $F$ be a quotient of $G$ constructed by Algorithm~2, with 
	$b_G(V_F^*)\leq D$ for $D\ge4k$, and $h\in V_F$ be
	an original singleton such that
	$d_F(v)\le D$ for all $v\in V_F\setminus\{h\}$. There is a randomized polynomial time algorithm,
	which we call later \textsc{UniqueHighDegreeSpanTree}$(F,h,D)$, that constructs
	a spanning tree $T_1$ of $F$ satisfying
	\begin{align}
		c_{F,T_1}(e)&=O(kDL^4)&&\text{for every }e\in E(T_1),\label{eq:local-edge}\\
		\ell_{F,T_1}(v)&=O(kDL^4)&&\text{for every }v\in V_F\setminus\{h\}.
		\label{eq:local-vertex}
	\end{align}
\end{lemma}
\begin{proof}
	If $V_F=\{h\}$, return the empty tree. Otherwise apply
	Theorem~\ref{thm:cancellation} in $F$ to $U=V_F\setminus\{h\}$,
	with $d=D$ and $b_0=D$ to get the partition $\CP$. 
	Since $F$ has at most $n$ vertices and
	total edge weight at most $m<n^2$, the partition
	$\CP$ satisfies
	\begin{align}
		b_F(A)&=O(DL^{3/2})\quad 		&& \forall A\in\calP,
		\label{eq:partition b}
		\\
		x_{\calP}(C)&\le O(L^{3/2})b_F(C) \quad && \forall C\subseteq U,\ b_F(C)\le D \ .
		\label{eq:partition x}
	\end{align}
	
	For $v\in U$, let $A(v)$ be the part $A\in \CP$ such that $v\in A$, and
	let 
	\[
	b(v)=\sum_{\substack{vw\in E_F\\w\notin A(v)}}\omega_F(vw)
	\]
	be the sum of weights of edges to neighbors of $v$ \emph{not} in 
	$A(v)$. Let $\graph=(F-h)/\CP$ be the quotient of $F-h$ defined by $\CP$,
	and define the \emph{supply} of every vertex $A\in \CP$ as $b(A)=\sum_{v\in A}b(v)$;
	we say that $b(v)$ is the \emph{contribution of $v$} to the supply $b(A)$ of $A$.
	Consider the following flow instance $I_{F,\CP}$ on $\graph$:
	Every vertex $A\in \CP$ is a source with supply $b(A)$, and the neighbors of $h$
	in $F_\CP$ are the sinks.
	Every edge has an infinite capacity, 
	and every vertex has capacity $c$ for a value $c>0$ 
	specified later; this capacitated graph is denoted $\graph^c$.
	The task is to find in $\graph^c$, if exists, a flow that 
	delivers all the supply to the sinks, subject to the capacity constraints.
	
	\begin{claim}\label{claim:flow}  
		There exists $c=O(kDL^3)$ such that the instance $I_{F,\CP}$ is feasible in $\graph^c$.
	\end{claim}
	On the high level,
	the proof of the claim goes along similar lines as the proof of the min-max connected
	cut Theorem~\ref{thm:repair}. The significant difference is that now we exploit the
	existence of a spanning tree of $G$ with congestion $k$; together with the partition
	of Theorem~\ref{thm:cancellation}, this makes it possible to get the flow. 
	\begin{proof}
		%
		For the analysis, let $R\in \CL$ be the original (vertex) region of the call, 
		that is, $F=G[R]/\CL(R)$ and $b_G(R)\le D$. 
		Recall the convention that contracting an isolating set $S_z$ retains the name $z$ 
		for the resulting vertex, and for every component $P$ of $T^*\setminus\{h\}$, 
		define $C_P=P\cap U$.
		We call $P$ a \emph{branch} (of $T^*$). Since $h$ is an original
		singleton, the sets $C_P$ partition the vertex set $U$, 
		and 
		by construction, for every branch $P$, 
		\begin{equation}\label{eq:branch-boundaries}
			b_{F}(C_P)\le b_G(P)\le k \ .
		\end{equation}
		
		Attribute the contribution $b(v)$ of the vertex 
		$v$ to the branch $P$ such that $v\in P$.
		The total contributions
		attributed to a branch $P$ is, exploiting~\eqref{eq:partition x}--\eqref{eq:branch-boundaries},
		\begin{equation}\label{eq:supply-family}
			\sum_{v\in C_P}\ b(v)
			\le2x_{\calP}(C_P)+b_F(C_P)=O(kL^{3/2}) \ .
		\end{equation}
		
		Call a branch $P$ $R$-\emph{external} if for some $v\in C_P$, the $v$--$h$ path in $T^*$ leaves $R$;
		all other branches are $R$-\emph{internal}.
		There are at most $b_G(R)\le D$ $R$-external
		branches: each contains a distinct tree edge leaving $R$.
		Define the \emph{projection of $P$ on $\CP$} as the set $P_\CP=\{A\in \CP : A^*\cap P\not = \emptyset \}$.
		
		For every $R$-internal branch $P$, for each $v\in C_P$, project the unique $v$--$h$ path in $T^*$
		to $F_{\CP}$, erase loops, and use the remaining path to route $b(v)$ units (which is the contribution
		of  $v$ to $P$) until the first vertex of $N_{F_{\CP}}(h)$.
		By the properties of $\CP$ and $R$, each part $A\in \CP$ intersects with projections of 
		at most $b_G(A^*)\le b_F(A)+b_G(R)=O(DL^{3/2})$ branches.
		
		For every $R$-external branch $P$, route the contribution of all vertices in $C_P$ along 
		arbitrary simple paths to $h$ in the connected graph $\graph$, stopping at the
		first neighbor of $h$. 
		
		We conclude that the total flow going in any $A\in \CP$ originates from at most
		$O(DL^{3/2})+D$ branches, therefore, by equation~\eqref{eq:supply-family}, its size
		is at most $O(kL^{3/2})(O(DL^{3/2})+D)=O(kDL^3)$, and this is also a bound
		on the load of any vertex other than $h$.
	\end{proof}
	
	The flow of Claim~\ref{claim:flow} can be obtained by repeated 
	max-flow computations: Let $l$ be a guess on the maximum load from 
	the claim (obtained, e.g., by trying all integer values for it, 
	smaller than $\sum_{A\in\calP}b_F(A)\le 2m$, starting with $1$, or by
	binary search on $[1,2m]$). 
	For each such guess, define
	for every node, other than $h$, its capacity as $l$ 
	and try to find a feasible solution (i.e., respecting the node capacities) 
	to the flow instance $I_{F,\CP}$.
	The flow corresponding to the smallest value of $l$ for which the instance
	is feasible, is the desired flow.
	Note that for the capacity $l=2m$, the instance is
	feasible; thus, the search always terminates.	
	
	Apply now the confluent-flow algorithm of Theorem~\ref{thm:confluent} on the flow
	from Claim~\ref{claim:flow}; the loss factor is
	$1+\ln|N_{\graph}(h)|=O(L)$. This gives us a confluent flow $f$ satisfying
	the instance $I_{F,\CP}$ with maximum node load $O(kDL^4)$; clearly,
	the same bound applies also to the maximum edge load. 
	Recall that the flow $f$ is realized by a collection of node disjoint trees.
	We combine them into a spanning tree $\tau$ of $\graph$ by adding all edges
	between the sinks $N_{\graph}(h)$ of $I_{F,\CP}$ and the vertex $h$.
	The construction of $\tau$ derived from the flow for the instance $I_{F,\CP}$
	ensures that the congestion of $\tau$ as well as the load of every vertex
	other than $h$ is $O(kDL^4)$.
	
	Every vertex in a part $A$ of $\CP$ has degree at most $D$ in $F$.
	Thus, for each $A\in \CP$, we can construct a spanning tree for $F[A]$
	by Lemma~\ref{lem:internal}; the bound on the edge and vertex loads of these local
	spanning trees is
	\[
	O\!\left(k(D+b_F(A))L^{3/2}\right)=O(kDL^3).
	\]
	Now substitute these ``local'' trees for the vertices of $\tau$, realizing tree edges
	of $\graph$ by edges of $F$ as follows:
	For each edge $AB$ of $\tau$, use any edge from $E_F(A,B)$.
	The result is a spanning tree of $F$.
	By Lemma~\ref{lem:expansion} and the above bounds, 
	the maximum edge load and vertex load for all vertices 
	but $h$, is at most $O(kDL^3)+O(kDL^4)=O(kDL^4)$. 
\end{proof}

At every call of Algorithm~2, Lemma~\ref{lem:internal} applies if
all degrees in $F$ are at most $D$. Otherwise the unique vertex
of degree greater than $D$ is an original singleton, and
Lemma~\ref{lem:local} applies to the call $(F,h,D)$. Thus the
chosen local tree has edge loads and vertex loads at all
non-singleton cells bounded by $O(kDL^4)$. This includes the root
and the leaves of the laminar family.

\subsection{Approximation Guarantee}

\begin{lemma}[Adding the costs of the levels]\label{lem:assembly}
	For every $D\ge4k$, the preceding construction returns, with high
	probability, a spanning tree $T$ of $G$ with
	\[
	c(G,T)=O(kDL^5).
	\]
\end{lemma}
\begin{proof}
	At every stage, the current bags partition $V_G$ and the current
	tree is a spanning tree of their weighted quotient. This follows
	inductively from Lemma~\ref{lem:expansion}. The local graph exposed
	when $A$ is expanded is exactly $G[A]/\CL(A)$, independently of
	how vertices outside $A$ are currently grouped. The local trees were constructed using only these induced graphs.
	The comparison with original cuts was needed for their analysis;
	the expansion accounting uses the current quotient.
	
	Track edge loads and vertex loads only at non-singleton bags.
	Each local tree bounds these two quantities by $O(kDL^4)$,
	at every call, including the root: use Lemma~\ref{lem:internal}
	when all local degrees are at most $D$, and Lemma~\ref{lem:local}
	otherwise.
	The high-degree vertex of Lemma~\ref{lem:local} is an
	original singleton, so it is never expanded and its vertex load need not
	be bounded.
	
	Suppose the two tracked quantities are bounded by $O(jkDL^4)$
	after $j$ processed levels. For an expanded bag $A$, the local load
	and its previous vertex load add in
	\eqref{eq:expand-edge}--\eqref{eq:expand-vertex}, giving
	$O((j+1)kDL^4)$ for new edges and new non-singleton cells. Surviving
	old loads do not increase. More explicitly, fix a common absolute
	constant for the local bounds; the same constant multiplied by $j$
	works in this induction. The family has $O(L)$ levels, so the final
	edge loads are $O(kDL^5)$.
\end{proof}

For the final approximation, compare the least-congested tree
obtained from Algorithm~2 over all tested thresholds with the
tree of congestion at most $8\sqrt{mn}$ from Proposition~\ref{prop:cci}.
Return the better tree.

\begin{theorem}\label{lem:approximation}
	There is a randomized polynomial-time algorithm with approximation
	ratio
	\begin{equation}\label{eq:ratio}
		O\!\left(\min\left\{kL^5,\frac{\sqrt{mn}}k\right\}\right)
		\le O(n^{2/3}L^{5/3})=\Ot(n^{2/3}).
	\end{equation}
\end{theorem}
\begin{proof}
	Some tested threshold satisfies $4k\le D<8k$, because $1\le k\le m$.
	Lemma~\ref{lem:family} guarantees progress for this threshold.
	All other guesses either return a family after at most $n-1$
	contractions or propagate \textsc{Fail}; the local routines also
	terminate in polynomial time for these guesses.
	Lemma~\ref{lem:assembly} gives a candidate of congestion $O(k^2L^5)$.
	Comparing it with the $8\sqrt{mn}$ candidate proves the first bound
	in~\eqref{eq:ratio}. As in Section~1, $m\le k(n-1)\le kn$.
	Consequently the ratio is at most
	\[
	O\!\left(\min\{kL^5,n/\sqrt{k}\}\right)
	\le O\!\left((kL^5)^{1/3}(n/\sqrt{k})^{2/3}\right)
	=O(n^{2/3}L^{5/3}).
	\]
	
	Each successful contraction contains at least two current vertices,
	so at most $n-1$ contractions occur for any guess. Every local graph
	has at most $n$ vertices and total weight at most $m$. The partition
	algorithm, node-capacitated flow, confluent rounding, and the
	hereditary-bisection routine are therefore polynomial-time operations
	also for guesses below $4k$; their congestion guarantees are needed
	only for a successful threshold.
	
	There are polynomially many randomized calls over all guesses.
	Amplify the bisection and quota calls using $O(\log n)$ independent
	repetitions, retaining the least-cost feasible output. A conditional
	union bound over the calls makes all required guarantees hold with
	probability at least $1-n^{-c}$, for any fixed $c>0$. The stopping
	argument in Theorem~\ref{thm:cancellation} then guarantees its
	inequalities simultaneously for all subsets, without testing them.
	Evaluating the congestion of each candidate ensures that other
	guesses cannot worsen the returned solution.
\end{proof}

\section{Hardness within \texorpdfstring{$2-\varepsilon$}{2-epsilon}}
\label{hard:section}\label{sec:hardness}
Next we prove the following hardness result. 
\begin{theorem}\label{hard:main}
	For every fixed $\varepsilon\in(0,1)$, there is no polynomial-time
	$(2-\varepsilon)$-approximation algorithm for spanning tree congestion,
	unless $\mathsf P=\mathsf{NP}$. 
\end{theorem}

The proof is by reduction from a restricted version of the vertex disjoint
path problem: given a graph $Q$ and $p$ pairs $(s_1,t_1),\ldots,(s_p,t_p)$ 
of vertices, can all $p$ pairs be joined by mutually vertex-disjoint paths? 
The problem is NP-complete even when the graph has maximum degree
three, all $2p$ terminals are distinct, and a vertex has degree one
if and only if it is a terminal~\cite[proof of Theorem~1]{hard:Iersel2018}.

We may also require $Q$ to be connected. First reject instances having
a pair in different components, and discard components without
terminals. In each remaining component subdivide an edge twice and
join consecutive components by an edge between distinct subdivision
vertices, forming a chain of components. Degrees remain at most three
and terminals retain degree one. A simple path between terminals in
one old component cannot leave that component: it would have to cross
a joining edge twice. The transformation therefore preserves the
answer. Instances with no terminal pair can be decided directly.
Write $N_Q=|V(Q)|$, $M_Q=|E(Q)|$, and $B=\max\{M_Q,2\}$.

Two observations explain the source restriction. Edge-disjoint paths
for distinct degree-one terminals in a graph of maximum degree three
are vertex-disjoint: a shared internal vertex would require four
incident edges, and a terminal cannot be internal to another path.
Conversely, the union of vertex-disjoint paths is a forest and can be
extended to a spanning tree of a connected graph.

For a positive-integer-weighted graph $H$, let
$c_w(H,R,e)$ be the total weight crossing the fundamental cut of
$e$ in a spanning tree $R$. Write $c_w(H,R)=\max_{e\in E(R)}c_w(H,R,e)$
and $\STC_w(H)=\min_R c_w(H,R)$.

Fix an integer $A\ge2$. Give the edges of $Q$ weight one. For every
pair $(s_i,t_i)$, add $A$ distinct vertices $w_{i1},\ldots,w_{iA}$ and
edges
\begin{equation}\label{hard:gadget}
	s_iw_{ij}\text{ of weight }1,\qquad
	w_{ij}t_i\text{ of weight }A+1
	\qquad(1\le j\le A).
\end{equation}
Call the resulting weighted graph $H_A$. The two-edge path
$s_iw_{ij}t_i$ is \emph{full} in a spanning tree if both its edges
belong to that tree. At most one such path for a fixed pair can be
full, since two would form a cycle.

\begin{lemma}\label{hard:gap}
	If the disjoint-path instance is a YES instance, then
	$\STC_w(H_A)\le A+B$. From any spanning tree of $H_A$ with congestion
	less than $2A$, one can extract a vertex-disjoint routing of all pairs
	in $Q$ in polynomial time. Consequently, a NO instance satisfies
	$\STC_w(H_A)\ge2A$.
\end{lemma}
\begin{proof}
	For a YES instance, extend its vertex-disjoint paths to a spanning
	tree $T_Q$ of $Q$, and attach every $w_{ij}$ to $t_i$ by its heavy
	edge. This gives a spanning tree $R$ of $H_A$. Consider
	$e\in E(T_Q)$ and the corresponding cut $(S,V(Q)\setminus S)$.
	All auxiliary vertices lie on the side of their $t_i$. Thus a gadget
	contributes $A$ if its terminals are separated and zero otherwise.
	At most one prescribed path uses $e$, and so
	\[
	c_w(H_A,R,e)
	=|\delta_Q(S)|+A\,|\{i:e\in E(P_{T_Q}(s_i,t_i))\}|
	\le M_Q+A.
	\]
	Every other tree edge isolates one auxiliary vertex and has load
	$1+(A+1)=A+2$. Hence $c_w(H_A,R)\le A+B$.
	
	Now let $R$ be any spanning tree of $H_A$ with congestion below $2A$.
	Suppose $s_iw_{ij}t_i$ is full. Deleting its heavy edge separates
	$s_i$ from $t_i$. The heavy edge contributes $A+1$ to this cut, and
	each of the other $A-1$ two-edge gadget paths contributes at least
	one. The heavy tree edge therefore has load at least $2A$, a
	contradiction. It follows that every auxiliary vertex is a leaf of
	$R$, and removing them leaves a spanning tree $T_Q$ of $Q$.
	
	If two prescribed paths in $T_Q$ used the same edge $e$, deleting
	$e$ in $R$ would separate both pairs. Each of the $A$ gadget paths
	for each pair must cross this cut and contributes at least one.
	Together with $e$ itself, the load would be at least $2A+1$.
	Consequently these paths are edge-disjoint, and the source restriction
	makes them vertex-disjoint. Removing auxiliary leaves and finding the
	paths takes polynomial time.
\end{proof}

Lemma~\ref{hard:gap} gives
\begin{equation}\label{hard:final-gap}
	\begin{array}{ll}
		\text{YES:}&\STC_w(H_A)\le A+B,\\
		\text{NO:}&\STC_w(H_A)\ge2A.
	\end{array}
\end{equation}

\begin{proof}[Proof of Theorem~\ref{hard:main}]
	Fix $\varepsilon\in(0,1)$ and take
	$A=\lceil2B/\varepsilon\rceil+1$. Then
	\[
	2A-(2-\varepsilon)(A+B)
	=\varepsilon A-(2-\varepsilon)B>0.
	\]
	A $(2-\varepsilon)$-approximation would return a tree of congestion
	below $2A$ on every YES instance, whereas no such tree exists on a
	NO instance. Evaluating the returned tree's congestion would decide
	the NP-complete source problem.
	To eliminate integer weights, we use the standard reduction 
	that replaces an edge with weight $w>1$ by $w$ internally disjoint 
	paths~\cite{LCH:25}.
	The construction is polynomial.
\end{proof}

\paragraph{AI Disclosure.}
The idea of using connected min-max multiway partition to approximate spanning tree congestion is from the authors,
specifically establishing a reduction when the graph has only one vertex with unbounded degree, with approximation
ratio dependent on the second largest degree $D$.
The main results in this paper were subsequently discovered by GPT-6 Astra; this includes the $\tilde{O}(n^{8/9})$ approximation algorithm,
the log-factor improvement in Proposition~\ref{prop:cci} combining \cite{Chandranetal:18,Hajiaghayietal:26}, and Theorems~\ref{thm:main}, \ref{thm:conminmaxcut}, \ref{hard:main}.

\appendix
\section{Omitted Proofs}\label{app:proofs}
\subsection{Proposition\ref{prop:cci}}\label{app:cci}
\begin{proofof}{Proposition~\ref{prop:cci}}
	The construction in~\cite[Theorem~10 and Algorithm~1]{Chandranetal:18}
	already gives this congestion. Its non-polynomial step is a connected
	partition with bounded degree sums. We show how to compute that partition.
	Let $H$ be an $\ell$-vertex-connected simple graph with $s$ vertices,
	$M$ edges, and distinct prescribed terminals $t_1,\ldots,t_\ell$.
	Set
	\[
	A=\left\lceil\frac{2M}{\ell}\right\rceil,\qquad
	w(v)=d_H(v)\quad(v\notin\{t_1,\ldots,t_\ell\}),\qquad
	c_i=A-d_H(t_i).
	\]
	Since $2M/\ell\ge s>\Delta(H)$, the capacities $c_i$ are nonnegative
	and their sum covers all nonterminal weights. Replace each edge by
	opposite arcs and delete outgoing terminal arcs. Menger's theorem
	gives $\ell$ paths from every nonterminal to distinct terminals,
	disjoint except at their starting vertex. Thus
	\cite[Theorem~3]{Hajiaghayietal:26} applies and computes connected
	parts $V_i$, each containing $t_i$ and no other terminal, with
	\[
	\sum_{v\in V_i}d_H(v)
	\le A+\max_{v\notin\{t_1,\ldots,t_\ell\}}w(v)-1
	\le \left\lceil\frac{2M}{\ell}\right\rceil+\Delta(H)-1
	\le \frac{4M}{\ell}.
	\]
	This is the degree-sum guarantee needed in line~15 of Algorithm~1
	in~\cite{Chandranetal:18}. Replacing that step preserves their
	congestion analysis and makes the whole construction polynomial-time.
\end{proofof}

\subsection{Lemma~\ref{lem:hereditary}}\label{app:hereditary}
\begin{proofof}{Lemma~\ref{lem:hereditary}}
	The edge-congestion bound is from the paper~\cite[Algorithm~1,
	Lemma~3.2, and Theorem~1.1]{Kolman2026}. Its hereditary bisection parameter is
	at most $B$: deleting edges from an induced subgraph cannot increase its
	minimum balanced-cut cost. Integer weights are represented by parallel unit
	edges.
	
	For the vertex bound, just note that each recursive cut has weight $O(B\sqrt{\log n})$, and
	its sides have at most three quarters of the current vertices. The recursive
	children are the connected components of the two sides. Every demand not
	internal to one child crosses this cut, so its total contribution to any
	vertex, including endpoints, is at most the cut weight. A fixed vertex lies on
	a chain of $O(\log n)$ recursive subproblems, ending at a singleton with no internal
	demand. Summing gives $O(B\log^{3/2}n)$.  
\end{proofof}

\subsection{Lemma~\ref{lem:internal}}\label{app:internal}
\begin{proofof}{Lemma~\ref{lem:internal}}
	For $X\subseteq C$, $|X|\ge2$, mark the vertices of $X$ in an 
	optimum spanning tree $T^*$ of $G$ and choose a centroid $x$:
	each component of $T^*-x$ has at most $|X|/2$ marks.
	If $x\in C^*$, select $v\in C$ such that $x\in\{v\}^*$.
	Apart from the possible component containing $v$, each marked
	component has a distinct first-exit tree edge from $\{v\}^*$
	toward a marked vertex outside this bag.
	There are at most
	$b_G(\{v\}^*)\le2D$ such edges.
	If $x\notin C^*$, each marked component has a distinct
	first-entrance tree edge into $C^\uparrow$, giving at most
	$b_G(C^*)\le b_F(C)+D$ marked components.
	
	Choose whole components carrying between $|X|/3$ and $2|X|/3$
	marks. A component with at least $|X|/3$ marks suffices; otherwise
	accumulate components until this threshold is reached. There are
	enough marks outside $x$, since $|X|-1\ge|X|/3$.
	Each component has original boundary at most $k$, so their union
	has boundary $O(k(D+b_F(C)))$. Intersecting this union with $X$
	gives a balanced cut of $F[X]$ with no greater cost. 
	Deleting edges only decreases this cost,
	so the same bound holds for the hereditary bisection of $F[C]$.
	Lemma~\ref{lem:hereditary} completes the proof.
\end{proofof}

\subsection{Proposition~\ref{prop:quota}}\label{app:quota}
\begin{proofof}{Proposition~\ref{prop:quota}}
	Consider the graph $Q$, $o\in V_Q$, $W$, the $\nu$-weight, $\ell$  and 
	as described in Proposition~\ref{prop:quota}. We modify $Q$ into a larger 
	$w$-weighted graph
	on which we apply the bisection approximation of Feige and Krauthgamer.
	Start with the graph $Q$ and add to it $W-\ell$ isolated vertices of 
	$w$-weight one and $W$ isolated vertices of $w$-weight zero. 
	Keep every edge of $Q$ with its capacity, give the vertex $o$ weight
	$w(o)=\ell$, and to every other vertex $v\in V_Q, v\not =o$, $w(v)=\nu(v)$.
	The total vertex weight is now $2W$.
	A weight-$W$ bisection side excluding $o$ contains vertices from $Q$ of 
	$w$-weight at least $W-(W-\ell)=\ell$; the $\nu$-weight of the set is the same. 
	Conversely, an admissible original
	set of weight $s\ge\ell$ can be extended with $W-s$ unit $w$-weight
	isolated vertices to obtain such a side. Cut costs and optimum values
	in both graphs are the same.
	
	Apply the edge-cost and integer-vertex-weight versions of the algorithm by
	Feige and Krauthgamer (Sections~5.1--5.2). The added zero-weight
	vertices ensure that the vertex weights are bounded by the extended
	order. This yields the claimed $O(\log^{3/2}(N+W))$ factor.
\end{proofof}

\section{Approximation of the connected min--max multiway cut}\label{app:mm}\label{app:conminmaxcut}
This appendix proves Theorem~\ref{thm:conminmaxcut}. We first show how
to make a terminal partition connected. The estimate keeps track of
the total boundary, which gives the edge-sensitive bound used in
the simpler STC algorithm.

\begin{theorem}[Connecting a terminal partition]\label{thm:repair}
	Let $F$ be a connected graph with positive integer edge weights,
	and let $\mathcal A$ be a partition into $t \ge2$ parts, each
	containing one prescribed terminal. Put
	\[
	B=\max_{A\in\mathcal A} b_F(A),\qquad
	S=\sum_{A\in\mathcal A} b_F(A).
	\]
	In polynomial time one can find a connected one-terminal partition
	with maximum boundary $O(\sqrt{SB}\log t)$.
\end{theorem}
\begin{proof}
	Split each initial part into its connected components and contract
	these components, obtaining a weighted graph $J$. Each vertex $x$
	of $J$ has degree $d_J(x)\le B$. The components of one initial part
	have no edges between them, so
	\[
	\sum_{x\in V(J)}d_J(x)=S\le tB.
	\]
	Exactly $t$ of the contracted vertices contain a terminal. Declare
	them sinks. Give every vertex $x$ the fixed supply $d_J(x)$, and
	put $\eta=\lceil\sqrt{S/B}\rceil$. Both $S$ and $B$ are positive.
	
	Give each edge capacity $\eta$ times its weight and try to route all
	supplies to the sinks. This is a single-commodity maximum-flow
	computation: a new source injects each supply, both orientations of
	an edge have its capacity, and each sink has an arc of capacity $S$
	to a new target. If all supply cannot be routed, a minimum cut gives
	a sink-free set $U_j$ in the current graph $J_j$ such that
	\begin{equation}\label{mm:deleted}
		\eta b_{J_j}(U_j)<\sum_{x\in U_j}d_J(x).
	\end{equation}
	Indeed, the source cut has capacity at most the total remaining
	supply, whereas cutting any sink-to-target arc costs $S$; a strictly
	violating set therefore contains no sink. Delete $U_j$ and repeat,
	always retaining the original supplies $d_J(x)$. At most $|V(J)|$
	iterations leave a feasible subgraph on a set $W$. Every sink
	remains, so $W$ is nonempty.
	
	Let $U$ be the union of the deleted sets. They are disjoint, and
	any edge from $U$ to $W$ was counted when its endpoint in $U$ was
	deleted. Summing~\eqref{mm:deleted} therefore gives
	\begin{equation}\label{mm:interface}
		\omega_J(U,W)\le S/\eta,
	\end{equation}
	where $\omega_J(U,W)$ is the total edge weight between $U$ and $W$.
	
	Stop the feasible flow at the first sink it reaches and cancel
	cycles. Incoming flow at $x$ is at most $\eta d_J(x)$, since the
	sum of the capacities of the incident edges is at most that
	quantity. Thus its node load, including its own supply, is at most
	$(\eta+1)B$. By the confluent-flow theorem in~\cite[Theorem~20]{Chen2007}, this flow can be converted
	in polynomial time into a flow in which all flow leaving a
	non-sink uses one outgoing edge, with node loads at most
	$(1+\ln t)(\eta+1)B$. The bound includes the loads at sinks and
	the supplies at all vertices. The positive supplies ensure that
	all vertices of $W$ belong to the resulting connected groups,
	each containing exactly one sink. The total supply in a group is
	its sink load, and it bounds that group's boundary in the original
	quotient $J$.
	
	Attach each whole connected component of $J[U]$ to one adjacent
	group. Such a neighbor in $W$ exists because $J$ is connected.
	There are no edges between different components of $J[U]$.
	Therefore, for any final group, the increase in its boundary is at
	most the entire interface $\omega_J(U,W)$, rather than the sum of the
	original degrees of the attached vertices. By~\eqref{mm:interface},
	the resulting maximum boundary is at most
	\[
	(1+\ln t)(\eta+1)B+S/\eta
	=O(\sqrt{SB}\log t).
	\]
	No deleted vertex contains a terminal. The final groups are
	connected and have one terminal each; expanding their constituent
	components gives the required partition of $F$.
\end{proof}

\begin{proof}[Proof of Theorem~\ref{thm:conminmaxcut}]
	Denote $B^* = OPT$.
	Use the min--max multiway-cut algorithm from ~\cite[Theorem~1.4]{BFK:14}. It produces a
	one-terminal partition, not necessarily connected, with maximum
	boundary
	\[
	B\le O\!\left(\sqrt{\log(n)\log(t)}\right)B^*
	\le O(\log n)B^*.
	\]
	An expected approximation guarantee suffices here: Markov's
	inequality gives constant success probability, and $O(\log n)$
	independent repetitions, selecting the smallest actual maximum
	boundary, give this bound with high probability.
	The optimum without connectivity is at most $B^*$, so it is a valid
	comparison value. For the returned partition,
	\[
	B\le S\le tB,\qquad S\le2 M.
	\]
	Apply Theorem~\ref{thm:repair} for $F = G$. Its maximum boundary is at most
	\[
	O\!\left(\min\{\sqrt t\,B,\sqrt{MB}\}\log(t)\right)
	=\Ot\!\left(
	\min\{\sqrt t\,B^*,\sqrt{MB^*}\}\right),
	\]
	as claimed. More explicitly, the two bounds are respectively
	\[
	O(\sqrt t\,B^*\log^2 n)
	\quad\text{and}\quad
	O(\sqrt{MB^*}\log^{3/2} n).
	\]
	For $t=1$ the whole connected graph is the optimal part, with zero
	boundary, and no approximation routine is needed.
\end{proof}

\newpage

\bibliographystyle{plainurl}
\bibliography{bibliography}
\end{document}